\documentclass[%
 reprint,
showpacs,%preprintnumbers,
 amsmath,amssymb,
 aps,prl
]{revtex4-1}

\usepackage{soul}
\usepackage{color}
\usepackage[usenames,dvipsnames]{xcolor}

\newcommand{\qt}{{\cal Q}}
\newcommand{\ft}{{\cal F}}

\usepackage{dcolumn}% Align table columns on decimal point
\usepackage{bm}% bold math
\usepackage{hyperref}% add hypertext capabilities
\begin{document}

\title{Quantum conversion}
\author{Michael Mazilu}
\email{michael.mazilu@st-andrews.ac.uk}
\affiliation{SUPA, School of Physics and Astronomy,\\ 
University of St Andrews, St Andrews KY16 9SS, UK
}%

\date{\today}

\begin{abstract}
%There are many ways to calculate the optical forces acting on scattering particles such as Maxwell's stress tensor, Lorentz forces, gradient and scattering forces, Lorenz-Mie formalism, T-matrix. All these approaches use the electromagnetic field and define the amount of linear momentum transferred to  scattering particles. 
The electromagnetic momentum transferred transfered to scattering particles is proportional to the intensity of the incident fields, however, the momentum of single photons  ($\hbar k$) does not naturally appear in these classical expressions. Here, we discuss an alternative to Maxwell's stress tensor that renders the classical electromagnetic field momentum compatible to the quantum mechanical one. This is achieved through the introduction of the quantum conversion which allows the transformation, including units, of the classical fields to wave-function equivalent fields. 
\end{abstract}


\maketitle

%\tableofcontents

\section{\label{sec:level1}Introduction}

Calculating the optical forces acting on mesoscopic particles typically relies on defining a balance equation between the momentum of the electromagnetic field and the momentum of the mesoscopic particle. This can be done using various definition of the electromagnetic stress tensor such as Chu, Maxwell-Lorentz, Minkowski, Abraham, Eistein-Laub\cite{Mansuripur:2013wn}. All these approaches deliver the same total momentum transfer to rigid particles. Further, in all these cases the momentum is proportional to the incident intensity. The differences between each stress tensor definition lies in their precise microscopic interpretation. In the following, we consider only Maxwell-Lorentz stress tensor, however all the discussions are applicable to any of the other definitions. 

%\section{Theory}

%\subsection{Quantum conversion}

To transform classical fields into quantum compatible wave-function fields, we introduce the quantum  conversion (QC). The QC is based on two parts, a mathematical transformation and a unit conversion. The transformation part corresponds to a convolution with the  Fourier transform of $1/\sqrt{\nu}$. On the other side, the unit conversion ensures the correct units in the physical context. The QC  applied to a function $f_0(t)$ with a  
zero-mean $\int^\infty_{-\infty}  f_0(t)dt=0$  defines the function $f(t)=(\qt\circ f_0)(t)$  as
%\begin{eqnarray}\label{eq:qmt}
%(\qt\circ f_0)(t) =\mathrm{p.v.} \int\limits_t^{\infty} \frac{f_0(\tau)}{\sqrt{h}} \frac{i-1}{\sqrt{t-\tau}} d\tau
%\end{eqnarray}
\begin{eqnarray}\label{eq:qmt}
(\qt\circ f_0)(t) =\mathrm{p.v.} \int\limits_{-\infty}^t \frac{f_0(\tau)}{\sqrt{h}} \frac{i-1}{\sqrt{t-\tau}} d\tau
\end{eqnarray}
where {\it p.v.} denotes the Cauchy principal value.  The integral operator defining the QC also includes Planck's constant, $h$, which converts between frequency and energy (intensity).  The linear transformation part of the QC operator is similar to the Hilbert transform  of $f_0(t)$ while normalising its spectral amplitude by $\sqrt{h \nu}$ where $\nu$ is the spectral frequency. Its fundamental property is summarised in the following Fourier space relationship between a function and its QC:
\begin{eqnarray}\label{eq:prop0}
\sqrt{h \nu}(\ft\circ f)(\nu)= (\ft\circ f_0)(\nu)
\end{eqnarray}
where $\ft$ corresponds to the Fourier transform in the time domain. %In words, this means that each spectral component of the signal $f_0(t)$ is 
Using this property and Perseval's formula it is possible to introduce an integral relationship between classical and quantum space
\begin{eqnarray}\label{eq:prop1}
\int_{-\infty}^\infty f^*_0(t)g_0(t) dt 
&=&\int_{-\infty}^\infty h \nu \hat f^*(\nu)\hat g(\nu) d\nu \nonumber\\
&=&\int_{-\infty}^\infty f^*(t) \left(i \hbar \frac{d}{dt} \right) g(t) dt 
\end{eqnarray}
which represents the main purpose of the QC. In effect, the QC introduces a first order time domain derivation whose role is to make the integral intensity measure proportional to the frequency of the field. As we will see in the following, this will introduce a proportionality between the energy/momentum measures and the frequency/wave-vector of the fields considered. The proportionality coefficient is Planck's constant and the QC converts using a different coefficients each spectral component of the field to encode the energy information onto the frequency of the field. This is possible as each frequency component is orthogonal to each other.  

For completeness, we can define the inverse QC as:  
\begin{eqnarray}\label{eq:iqmt}
f_0(t)=(\qt^{-1}\circ f)(t)& =& i \hbar \frac{d}{dt} (\qt\circ f)(t)
\end{eqnarray}
and exemplify the effect of the QC on a  harmonic function $\qt(\exp(i\omega \tau))(t)=\mathrm{sgn} (\omega) \; i \exp(i \omega t)/\sqrt{\hbar \omega }$ where $\mathrm{sgn}(w)$ corresponds to the sign function. 

%\subsection{Application}

In the following, we consider the QC of the electromagnetic fields solutions of Maxwell's equations (Gaussian units) in free space,
\begin{eqnarray}
\nabla\cdot\mathbf{E}_0 & =& 0,\cr
\nabla\cdot\mathbf{B}_0 =&  0,\cr
c\nabla\times\mathbf{E}_0  &= & -\partial_{t}\mathbf{B}_0,\cr
c\nabla\times\mathbf{B}_0 & =& \partial_{t}\mathbf{E}_0,
\label{Eq:Maxwell}
\end{eqnarray}
where $\mathbf{E}_0$ and $\mathbf{B}_0$ are the electric and magnetic vector fields  and where $c$ is the speed of light. We define the QC of the electromagnetic fields as $\mathbf{E}=\qt\circ\mathbf{E}_0$ and $\mathbf{B}=\qt\circ\mathbf{B}_0$ where $\mathbf{E}_0$ and $\mathbf{B}_0$ are considered to be zero mean analytic signal functions i.e. contain only strictly positive frequencies.

Taking property (\ref{eq:prop1}) into account we can define the converted field energy density as 
\begin{eqnarray}
\rho_{\cal E}=\frac{1}{8 \pi}(\mathbf{E}^* (i\hbar \partial_t) \mathbf{E}+\mathbf{H}^* (i\hbar \partial_t) \mathbf{H})
\end{eqnarray} 
which will is globally identical to the standard definition of energy density i.e. the total time integrated energy in a region is the same for both definitions. The time dependence of the fields has change however using the normalised fields gives exactly the same total intensity/energy of the pulse. What is more this approach maintains relative intensities which means that through relative measures we are not able to distinguish physically between the classic and the quantum version. 
The origin of property is the orthogonality of the different spectral components of the pulse i.e. the total energy content of two monochromatic waves is equal to the sum of their individual energies.

The flow of energy is described by  
\begin{eqnarray}
\mathbf{j}_{\cal E}=\frac{c}{8 \pi}\left(\mathbf{E}^* \times (i\hbar \partial_t) \mathbf{H}+(i\hbar \partial_t) \mathbf{E} \times  \mathbf{H}^*\right).
\end{eqnarray} 
Using this definition of the flow allows us to determine the energy transferred to a volume $V$ surrounded by a surface $S$ as
\begin{eqnarray}\label{eq:int}
\frac{d{\cal E}}{dt}=\int_{S}\mathbf{j}_{\cal E}\cdot \mathbf{n} \;ds  
\end{eqnarray} 
where $\mathbf n$ is the surface normal and ${\cal E}=\int_V \rho_{\cal E}\; dv$ the integrated energy density in the volume $v$.

Using the quantum converted fields we can redefine the linear momentum density and its current/flux. To do this, we notice the parallelism between the definition of energy and momentum. They only differ in the generating differential operator. What is more, taking into account the dispersion relationship $ck=\omega$ we can define the density of the linear momentum as :
\begin{eqnarray}
\rho_{Pi}=\frac{1}{8 \pi}(\mathbf{E}^* ( P_i ) \mathbf{E}+\mathbf{H}^* ( P_i ) \mathbf{H})
\end{eqnarray} 
where $P_i$ is a place holder for differential operators $(i\hbar)\partial_x$, $(i\hbar)\partial_y$ and $(i\hbar)\partial_z$ corresponding respectively to the linear momentum  in the $x-$, $y-$ and $z-$direction. 

The conservation relationship implies that the transfer of momentum (force acting on scattering body) can be calculated by integrating the momentum flux through a surface surrounding the scattering object. Replacing, in equation (\ref{eq:int}), the energy operator with a momentum operator  delivers the variation of the momentum in the chosen direction. This momentum variation acts as a force on the scattering object. For example, in the continuous wave case we can calculate the optical period averaged force acting of the object 
\begin{eqnarray}\label{eq:int}
F_i= \frac{c}{8 \pi}\int_{S}\left(\mathbf{E}^* \times (P_i) \mathbf{H}+(P_i) \mathbf{E} \times  \mathbf{H}^*\right)\cdot \mathbf{n} \;ds.  
\end{eqnarray} 
This integral delivers identical average forces as Maxwell's stress tensor. It shows the equivalence between quantum converted fields in conjunction with the operator approach and classical electromagnetic forces.  In both cases, the observable macroscopic mechanical force is the same. 

We further remark that the density defined by the identity operator $\rho=(\mathbf{E}^*\mathbf{E}+\mathbf{H}^*  \mathbf{H})/(8\pi)$  corresponds to the photon density. This explains physically the equivalence between the momentum flux integral (\ref{eq:int}) and Maxwell's stress tensor, which in this context determines the variation of photon flux due to a scattering body.

To complete the equivalence picture between the classical fields and the quantum fields we further need to take into account the optical eigenmodes defined either over the whole space \cite{Mazilu:2011gx,Mazilu:2009dy} or over a finite region of interest \cite{Mazilu:2011uf}. Indeed, using this approach one can draw a direct equivalence between   quantum mechanics and classical electromagnetism where the optical eigenmodes play the role of photon wavefunctions \cite{BialynickiBirula:2003ub}.

Further, the quantum conversion can easily be modified to deal with non analytic functions and negative frequencies when considering relative energy transfers. In physical terms, the quantum conversion corresponds to the introduction of frequency dependent units or measurement metric. This conversion generalises the concept of position dependent metric to the  Hilbert space. Indeed, each orthogonal basis vector/function can have its own units just like each point in space can have its own distance measure. Finally, we note that the inverse quantum conversion can be applied to transform quantum wave functions into their classical form opening up to the possibility of a new interpretation of quantum mechanics.

%\bibliography{qc.bib} 
%\bibliographystyle{spiebib}   

\end{document}